\documentclass[preprint,5p,times,twocolumn]{elsarticle}

\usepackage{amssymb}
\usepackage{amsmath}
\usepackage{amsthm}

\usepackage{multirow}
\usepackage{caption}
\usepackage{subcaption}

\usepackage{subeqnarray}
\usepackage{cases}

\usepackage[linesnumbered,vlined,ruled,boxed]{algorithm2e}
\usepackage{xcolor}
\newcommand{\bluecomment}[1]{\textcolor{blue}{#1}}
\usepackage{float}
\usepackage{hyperref}

\begin{document}

\begin{frontmatter}



\title{GAWM: Global-Aware World Model for Multi-Agent Reinforcement Learning}


\author[1]{Zifeng Shi}
\ead{shizifeng@zju.edu.cn}

\author[2,1]{Meiqin Liu}
\ead{liumeiqin@zju.edu.cn}

\author[1]{Senlin Zhang}
\ead{slzhang@zju.edu.cn}

\author[1]{Ronghao Zheng}
\ead{rzheng@zju.edu.cn}

\author[1]{Shanling Dong}
\ead{shanlingdong28@zju.edu.cn}

\author[2]{Ping Wei}
\ead{pingwei@mail.xjtu.edu.cn}

\affiliation[1]{
	organization={College of Electrical Engineering},
	addressline={Zhejiang University}, 
	city={Hangzhou},
	postcode={310027}, 
	country={China}
}

\affiliation[2]{	
	organization={National Key Laboratory of Human-Machine Hybrid Augmented Intelligence},
	addressline={Xi'an Jiaotong University}, 
	city={Xi'an},
	postcode={710049}, 
	country={China}
	}


\begin{abstract}
In recent years, Model-based Multi-Agent Reinforcement Learning (MARL) has demonstrated significant advantages over model-free methods in terms of sample efficiency by using independent environment dynamics world models for data sample augmentation. However, without considering the limited sample size, these methods still lag behind model-free methods in terms of final convergence performance and stability. This is primarily due to the world model's insufficient and unstable representation of global states in partially observable environments. This limitation hampers the ability to ensure global consistency in the data samples and results in a time-varying and unstable distribution mismatch between the pseudo data samples generated by the world model and the real samples. This issue becomes particularly pronounced in more complex multi-agent environments. To address this challenge, we propose a model-based MARL method called GAWM, which enhances the centralized world model's ability to achieve globally unified and accurate representation of state information while adhering to the CTDE paradigm. GAWM uniquely leverages an additional Transformer architecture to fuse local observation information from different agents, thereby improving its ability to extract and represent global state information. This enhancement not only improves sample efficiency but also enhances training stability, leading to superior convergence performance, particularly in complex and challenging multi-agent environments. This advancement enables model-based methods to be effectively applied to more complex multi-agent environments. Experimental results demonstrate that GAWM outperforms various model-free and model-based approaches, achieving exceptional performance in the challenging domains of SMAC.
\end{abstract}








\end{frontmatter}



\section{Introduction}

Multi-Agent Reinforcement Learning (MARL) offers a flexible and powerful approach to decision-making in environments involving multiple agents. By optimizing the coordination of agent interactions, MARL has been successfully applied to various tasks requiring both cooperative and competitive strategies, such as multi-agent games~\cite{rashid2020monotonic,Baker_Kanitscheider_Markov_Wu_Powell_McGrew_Mordatch_2019,ye2020mastering}, multi-agent cluster control~\cite{Matignon_Jeanpierre_Mouaddib_2022,Hung_Givigi_2017,RAMEZANLOU2024129005}, and autonomous driving~\cite{You_Lu_Filev_Tsiotras_2019,Shalev-Shwartz_Shammah_Shashua_2016, GAO2024128482}. However, due to the partial observability and high dimensionality of observation information, as well as the non-stationarity caused by multi-agent cooperative strategy optimization, a large amount of environmental interaction data is required to ensure policy convergence. In real-world scenarios, the resources and time required to collect such data are often prohibitive. This highlights the critical importance of sample efficiency.

To address this issue, Model-based Reinforcement Learning (MBRL) generates pseudo data samples by constructing models of environment interaction dynamics, thus reducing the reliance on large quantities of real data samples. To further improve the world model's ability to represent agent state features, latent-variable-based world models have been introduced, achieving significant success in single-agent settings~\cite{Hafner_Lillicrap_Ba_Norouzi_2019,Jänner_Fu_Zhang_Levine_2019, Moerland_Broekens_Plaat_Jonker_2023, MALEKZADEH2023165}. Moreover, by aligning the consistency between global information from the world model and local agent-specific observations, this approach has been extended to Multi-Agent Reinforcement Learning (MARL)~\cite{Krupnik_Mordatch_Tamar_2019,Egorov_Shpilman_2022,Wu_Yu_Chen_Hao_Zhuo_2023, MACD}. However, the accuracy constraints of the world model in capturing the dynamics of environmental interactions significantly impact the reliability of sample trajectory generation. This hinders the diversified exploration of the real trajectory sample space, making the effective prediction space of the world model narrow and inaccurate~\cite{Wu_Yu_Chen_Hao_Zhuo_2023}. Specifically, the performance of these methods is still limited by several key issues that prevent them from fully realizing their potential.

\begin{figure}[h]
	\centering
	\includegraphics[width=\linewidth]{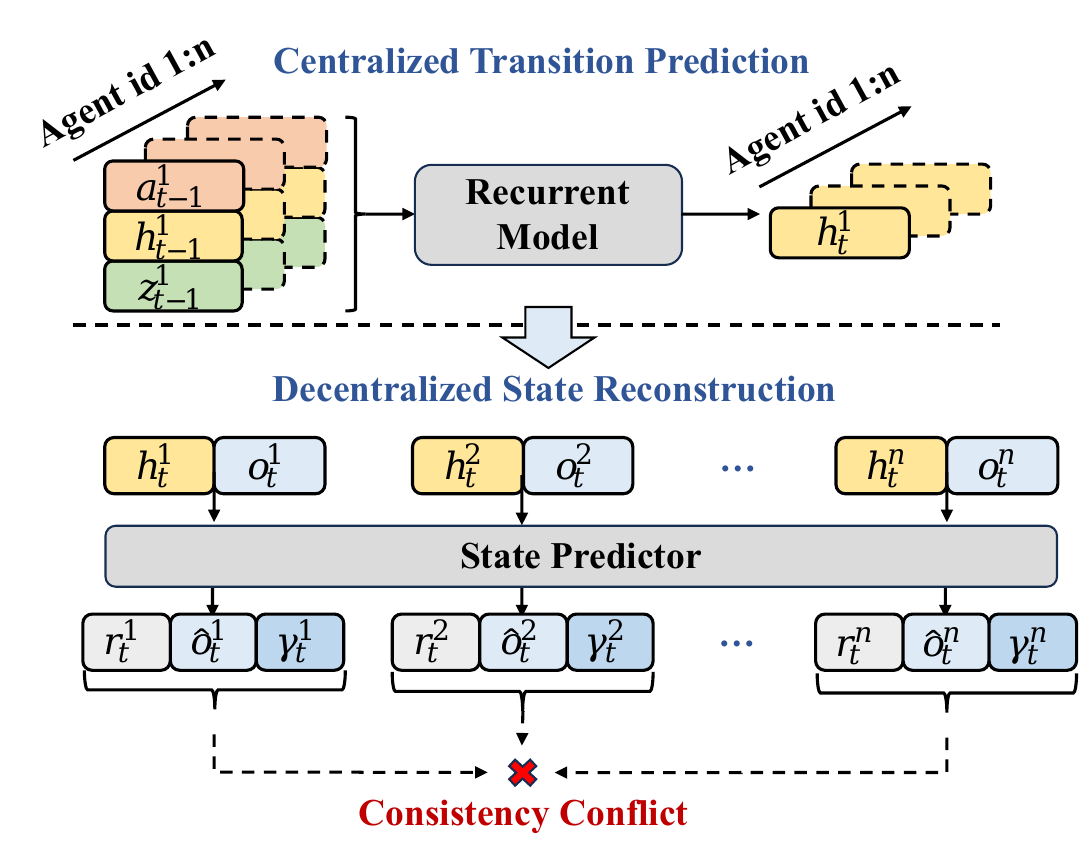}
	\caption{The current mainstream world models adopt a centralized state-transition prediction and distributed state-reconstruction framework. In this approach, the inputs for state-transition prediction include global latent state variables and action information, while the current state representation and reconstruction rely solely on locally observable state information. Due to the inherent limitations of partial observability, each agent's local observations provide only a fragmented view of the global state, making it difficult to accurately predict and represent global information. Consequently, this limitation may lead to inconsistencies in the reconstructed state information (e.g., rewards, observations, and discount factors) and cause conflicts in global consistency.}
	\label{fig:pro}
\end{figure}

Firstly, as shown in Fig.\ref{fig:pro}, existing world models\cite{Egorov_Shpilman_2022,Wu_Yu_Chen_Hao_Zhuo_2023} predominantly adopt a centralized state-transition prediction and decentralized state-reconstruction paradigm. For each agent $i,i\in[1,n]$ of $n$ agents, this approach relies solely on the current local observation ($o_t^i$ ) of individual agent $i$ and global historical latent state ($\boldsymbol{h}_t$) to represent each agent's current lantent state ($z_t^i$ ), without a unified fusion of instantaneous local observations across agents. In this case, it will be extremely difficult to reconstruct accurate global information of multi-agent systems based on $z_t^i$. Moreover, these models struggle to ensure global consistency of partial observations in partially observable environments. The lack of global consistency in local information may lead to contradictions in the predicted global state information, including reconstructed team rewards, discount factors, and local observation data. Such inconsistencies can lead to conflicting convergence directions, heightened instability during optimization, and ultimately diminished final performance. In fact, since the world model itself is trained in a centralized manner, decentralized state reconstruction are not necessary.

Secondly, due to the fact that the training sample data for the world model comes from the interaction exploration between the agent and the environment, its data sample distribution is always in a dynamic process of change. This results in the online dynamic learning process of the world model also undergoing dynamic changes. When the distribution of data samples changes dramatically, this may result in the generation of data samples that also deviate significantly from the true sample distribution. This instability leads to unreliable pseudo-sample generation, which can disrupt the training process and hinder the agent’s ability to learn effective policies. 

Lastly, previous approaches directly use the representation vectors produced by the world model as inputs to the policy network, leading to a lack of decoupling between the world model and the policy model. This integration either conflicts with the centralized training, decentralized execution (CTDE) paradigm or incurs significant computational costs, limiting the scalability and practical deployment of these methods in real-world multi-agent systems.

In this work, we propose a global-aware world model for MARL, called GAWM. GAWM offers three key contributions.
\begin{itemize}
	\item \textit{Local Observation Fusion Representation.} GAWM introduces a multi-agent world model that effectively integrates the local observation information from different agents for state representation, thereby substantially enhancing the global consistency of multi-agent state representations in complex environments. 
	\item \textit{Team Reward Trend Modeling.} GAWM adopts trend modeling for team rewards instead of precise modeling, which reduces reward modeling complexity and enhances the robustness of online world model learning without impacting policy convergence.
	\item \textit{CTDE Paradigm.} Unlike previous model-based CTCE approaches ~\cite{Egorov_Shpilman_2022, Wu_Yu_Chen_Hao_Zhuo_2023},  GAWM decouples the world model from the policy model, fully implementing a concise and lightweight CTDE paradigm in standard scenarios.
\end{itemize}     
Experimental results on various tasks in the StarCraftII~\cite{Samvelyan_Rashid_Witt_Farquhar_Nardelli_Rudner_Hung_Torr_Foerster_Whiteson_2019} benchmark show that GAWM consistently outperforms the existing methods.

\section{Related works and Preliminaries}
\subsection{MARL}
In most Multi-Agent Reinforcement Learning (MARL) problems, the process is defined as a Decentralized Partially Observable Markov Decision Process (Dec-POMDP)~\cite{Oliehoek_Amato_2016}. This is represented by the tuple $\langle N, S, A, P, R, \gamma, \Omega, O \rangle$, where $N$ is the number of agents, $S$ is the global state space, and $A = \prod A^i$ is the joint action space of all agents. The state transition is governed by the probability function $P(\boldsymbol{s}_{t+1}|\boldsymbol{s}_t, \boldsymbol{a}_t)$, and $R(\boldsymbol{s}_t, \boldsymbol{a}_t)$ denotes the reward function based on the joint action $\boldsymbol{a}_t=\{a^1, ..., a^n | a^i \in A^i, i \in \{1, ..., n\}\}$ in state $\boldsymbol{s}_t \in S$. The discount factor $\gamma \in (0,1]$ defines the significance of future rewards. The observation space is represented by $\Omega(s)$, and the observation mapping function $O(s^i)$ defines the partial observation $o_t^i$ that agent $i$ receives for state $s_t^i$. At each timestep $t$, agent $i$ selects an action $a^i_t$ based on the policy $\pi_i(a^i_t|\tau^i_t)$, where $\tau^i_t$ is the history of actions and observations. The environment then returns the team reward $r_t = R(\boldsymbol{s}_t, \boldsymbol{a}_t)$, and the global state $s_t$ evolves according to the transition function $P(\boldsymbol{s}_{t+1}|\boldsymbol{s}_t, \boldsymbol{a}_t)$. The objective in MARL is to maximize the expected return of the joint policy $\boldsymbol{\pi} = J(\pi^1, ..., \pi^n) := \mathbb{E}_{\pi}[\sum_{t'=0}^{\infty} \gamma_{t'} r_{t+t'} | \boldsymbol{s}_t, \boldsymbol{a}_t]$.
\subsection{Single-agent MBRL}
To address the high sampling cost, Model-Based Reinforcement Learning (MBRL) uses self-supervised learning to build an interactive dynamics model, known as the world model, which estimates the state transition probability distribution and reward function. It has been shown that utilizing the world model to expand the sample set improves sample efficiency~\cite{Jänner_Fu_Zhang_Levine_2019, Feinberg_Wan_Stoica_Jordan_Gonzalez_Levine_2018, Ayoub_Jia_Szepesvári_Wang_Yang_2020}. Given the complexity of dynamic interactions in high-dimensional environments, latent variable world models have been proposed to represent state transitions in such scenarios. For example, the Dreamer series~\cite{Hafner_Lillicrap_Ba_Norouzi_2019, Hafner_Lillicrap_Norouzi_Ba_2020, hafner2023dreamerv3} uses the Recurrent State Space Model (RSSM), while methods like IRIS~\cite{Micheli_Alonso_Fleuret_2022}, Storm~\cite{Zhang_Wang_Sun_Yuan_Huang_2023}, and TWM~\cite{Robine} employ Transformers~\cite{Vaswani_Shazeer_Parmar_Uszkoreit_Jones_Gomez_Kaiser_Polosukhin_2017} to update their latent states. These approaches map current state information into a latent space, recursively estimate the next latent state, and then reconstruct the state information back into the original low-dimensional space. This temporal process allows for the simulation of agent-environment interactions and the generation of pseudo trajectories, thereby improving sample efficiency.
\subsection{Multi-agent MBRL}
With the widespread application of latent variable world models, world models have gradually begun to be applied to the multi-agent paradigm. MAMBA~\cite{Egorov_Shpilman_2022}, as a pioneering model-based MARL effort inspired by DreamerV2~\cite{Hafner_Lillicrap_Norouzi_Ba_2020}, introduced a world model specifically designed for multi-agent environments. Building on MAMBA, MAG~\cite{Wu_Yu_Chen_Hao_Zhuo_2023} addressed the issue of local model prediction errors propagating through multi-step rollouts by treating local models as decision-making agents, significantly improving prediction accuracy in complex multi-agent environments. Although MAMBA and MAG demonstrate improvements in sample efficiency compared to model-free methods, their applicability is constrained by the CTCE paradigm, and there remains considerable potential for further enhancement in their asymptotic convergence performance. To implement the CTDE paradigm, MACD~\cite{MACD} employs a two-level latent variable world model. The upper-level global model learns the global latent states, while the lower-level local model takes the global latent state features from the upper-level model to predict local states. During the inference phase, agents only need to use the lower-level local model for reasoning, enabling decentralized execution of the learned policy. However, this approach requires assigning a world model to each agent. As the number of agents increases, it lacks a unified fusion of the local observations across all agents, which limits its performance in complex environments. Additionally, the method introduces supplementary global state information, thereby increasing the demand for extensive information processing and reliance on global states. In contrast, GAWM not only adheres to the CTDE paradigm but also enhances the world model's ability to represent global states, significantly improving convergence performance.

\section{Methodology}
\label{sec:theory}
We propose the Global aware world model (GAWM) method, which is a novel model-based MARL algorithm that adopts a latent variable world model architecture. What sets our world model apart from previous work is its ability to ensure global consistency and stability in data sample generation, thereby enabling the CTDE policy to converge more stably in relatively complex high-dimensional environments. Specifically, without introducing additional global information, GAWM significantly enhances the global representation ability of latent variables for the current multi-agent state by adopting global-aware state transition prediction and reconstruction. In addition, modeling the trends of team rewards significantly reduces the complexity of finely characterizing team rewards within the world model, while ensuring that policy convergence remains unaffected. This approach not only enhances the world model's ability to effectively capture the overall trends in team rewards but also makes its training process more stable. In this section, we first describe our novel world model architecture and introduce how it significantly increases the global consistency of data sample generation and the temporal stability of sample distribution. Then we provided a detailed introduction on how we implemented the MARL algorithm for CTDE.
\subsection{Architecture}
\label{sec:arch}
The architecture of GAWM, as shown in Eq.(\ref{for:RSSM}) and  Eq.(\ref{form:pre}), includes RSSM models and predictors. GAWM not only uses action fusion for temporal state prediction (as shown in Eq.~(\ref{eq:actfus})), but also introduces an additional block of observation fusion (as shown in Eq.~(\ref{eq:obsfus})) to further facilitate the integration of local observations among the agents, thus enhancing the global characterization of the current latent state.
\begin{subequations}
\begin{numcases}{\text{RSSM}}
	\text{Recurrent model:}   & $h_t^i = f_{rec}(h_{t-1}^i, e^i_t)$, \label{eq:rec}\\
	\text{Act-fusion:}   & $e^i_t = f^i_{af}(\boldsymbol{z}_t, \boldsymbol{a}_t)$, \label{eq:actfus}\\
	\text{Obs-fusion:}   & $g^i_t = f^i_{of}(\boldsymbol{h}_t, \boldsymbol{o}_t)$, \label{eq:obsfus}\\
	\text{Posterior model:}   & $z^i_t \sim p_{post}(z^i_t \mid g^i_t)$, \label{eq:posterior}\\
	\text{Prior model:}     & $\hat{z}^i_t \sim p_{piror}(\hat{z}^i_t \mid h^i_t)$, \label{eq:prior}
\end{numcases}
\label{for:RSSM}
\end{subequations}
\begin{subequations}
\begin{numcases}{\text{Predictors}}
\text{Observation:}        & $\hat{o}^i_t \sim p_{obs}(\hat{o}^i_t \mid h^i_t, z^i_t)$,\\
\text{Reward:}      &$\hat{r}_t \sim p_{rew}(\hat{r}_t \mid \boldsymbol{h}_t, \boldsymbol{z}_t)$,\\
\text{Discount:}    &$\hat{\gamma}_t \sim p_{dis}(\hat{\gamma}_t \mid \boldsymbol{h}_t, \boldsymbol{z}_t)$.
\end{numcases}
\label{form:pre}
\end{subequations}

\subsubsection{Global-aware World Model}
\paragraph{Recurrent Model} The recurrent model, illustrated in Eq.~(\ref{eq:rec}), employs a GRU~\cite{GRU} structure to capture environmental dynamics in partially observable multi-agent scenarios. It integrates historical and current state information using deterministic embeddings \( h_t \) and stochastic embeddings \( z_t \).
\paragraph{Act-Fusion}
Similar to other multi-agent MBRL approaches, GAWM's Act-Fusion module leverages Transformers~\cite{Vaswani_Shazeer_Parmar_Uszkoreit_Jones_Gomez_Kaiser_Polosukhin_2017}. In multi-agent systems, interactions between agents often involve diverse actions with significant global complexity. This model captures global interaction features by fusing cross-agent action information. During the fusion process, stochastic embeddings \( \boldsymbol{z}_t \) and actions \( \boldsymbol{a}_t \) interact across agents, generating the global-action-aware input embeddings \(e_t^i\), which are essential for the recurrent model to update the historical state \(h_t\).  
\paragraph{Obs-Fusion}
Unlike other previous works, GAWM has a novel obs-fusion model. Considering that the amount of information contained in the current latent state \( z_t^i \) constructed directly from local observations is insufficient to reconstruct an accurate global state information, we design an information fusion model that enhances the global information representation. This model takes local observation \( o_t^i \) and historical latent states \( h_t^i \) as inputs, and uses Transformers~\cite{Vaswani_Shazeer_Parmar_Uszkoreit_Jones_Gomez_Kaiser_Polosukhin_2017} to achieve cross agent information extraction and fusion, outputting accurate and globally consistent current latent state \( z_t^i \).
\paragraph{Posterior Model} The posterior model, described in Eq.~(\ref{eq:posterior}), predicts \(\boldsymbol{z}_t\) given the observation \(\boldsymbol{o}_t\), providing a foundation for reconstructing other variables. This task is simplified by minimizing the evidence lower bound~\cite{Kingma_Welling_2013}. Unlike previous work, GAWM utilizes the output of obs-fusion model as the input for the posterior model, enabling it to integrate state information from multiple agents for enhanced representation.
\paragraph{Prior Model} The goal of the prior model is to predict $z_t^i$ as accurately as possible without prior information $o_t^i$, as shown in Eq.~(\ref{eq:prior}). It is trained by minimizing the Kullback-Leibler (KL) divergence between \(\hat{z}^i_t\) and \({z}^i_t\) to approximate the posterior model. Thus, the world model can forecast future trajectories without the true observation information and generate samples for training the policy model.
\paragraph{Reconstruct Predictors} As shown in Eq.~(\ref{form:pre}), observation, reward and discount predictors are employed to reconstruct \(\boldsymbol{o}_{t}\), \(r_{t+1}\),  and \(\gamma_{t}\) from \(\boldsymbol{h}_{t}\) and \(\boldsymbol{z}_{t}\). Unlike previous work, we also consider the issue of global consistency in the design of the global information predictor. When predicting global state information such as team rewards and discount factors, we directly use the potential state information of all agents as input for prediction. This will further ensure the consistent representation of our algorithm on global information. These predictors are trained via supervised loss. The world model joint loss includes temporal prediction KL divergence loss and predictor reconstruction loss. Minimize the joint loss function through gradient descent to update the world model. 
\begin{align}
\label{align:jointlossfunc}
\begin{split}
\mathcal{L}_{\mathcal{M}}({\theta_{\mathcal{M}}})&=\sum_{t=1}^{T}-\ln p\left(\hat{\boldsymbol{o}}_{t} \mid \boldsymbol{h}_{t}, \boldsymbol{z}_{t}\right)-\ln p\left(\hat{{r}}_{t} \mid \boldsymbol{h}_{t}, \boldsymbol{z}_{t}\right) \\& \quad -\ln p\left(\hat{\gamma}_{t} \mid \boldsymbol{h}_{t}, \boldsymbol{z}_{t}\right)+  \beta\mathcal{L}_\mathcal{KL}\left[\boldsymbol{z}_{t}| | \hat{\boldsymbol{z}}_{t}\right] \\
&=\mathcal{L}_{rec}({\theta_{\mathcal{M}}}) + \beta\mathcal{L}_\mathcal{KL}({\theta_{\mathcal{M}}}).
\end{split}
\end{align}
\paragraph{Reward Trend Modeling}
\label{sec:imp}
MBRL generates pseudo trajectories with predicted rewards to train policies. However, accurately modeling rewards is challenging due to the dynamic complexity of environment interactions. Significant reward bias can severely impact the convergence process of the policy \(\pi\). Inspired by DreamSmooth~\cite{lee2023dreamsmooth}, we replace precise reward predictions with approximate estimates in environments characterized by high complexity and sparse rewards. Given the similarities in MARL environments, GAWM incorporates temporal smoothing of team rewards within each episode while maintaining total reward consistency:  
\begin{align}
\label{rewardSmooth}
\hat{r}_t\leftarrow f(r_{t-H:t+H}) &= \sum_{i=-H}^Hf_i\cdot r_{clip(t+i,0,T)} \quad  \textnormal{s.t.}  \sum_{i=-H}^{H}f_i=1,\\
\label{formula:rewardsmoothing}
f_i &= \frac{\exp\left(-\frac{i^2}{2\sigma^2}\right)}{\sum_{i=-H}^{H} \exp\left(-\frac{i^2}{2\sigma^2}\right)},
\end{align}
where \(T\) and \(H\) represent the episode horizon and smoothing window, respectively. This approach smooths reward data over time, using the processed rewards to train the reward model, allowing it to better fit the smoothed reward distribution. In our experiments, Gaussian smoothing was applied to the reward function, as defined in Eq.~(\ref{formula:rewardsmoothing}). Importantly, using smoothed rewards in MARL does not compromise strategy optimality.
\subsubsection{CTDE Policy} 
\label{policy}
TMost existing model-based methods use the centralized feature representations \(\boldsymbol{h}_t\) and \(\boldsymbol{z}_t\) from the world model as input for the policy model during both training and execution. In contrast, the policy model \(\boldsymbol{\pi}\) in GAWM directly takes the distributed, local observations \({o}_t^i\) (where \(i \in \{1, \dots, N\}\)) of each agent as input during both training and execution. During training, these local observations are reconstructed using the centralized world model, whereas during execution, they are directly acquired by agents interacting with the environment. Additionally, GAWM integrates GRU units into the policy model to better leverage historical information. The policy model is then used to compute each agent’s action distribution as \({a}_t^i\sim{\pi}^i({a}_t^i|{o}_t^i)\). By decoupling the world model from the policy model, this architecture ensures that GAWM adheres to the CTDE paradigm in standard scenarios.

GAWM employs the MAPPO method \cite{yu2022surprising} for its policy model \(\boldsymbol{\pi}\), leveraging an Actor-Critic architecture. The Actor (policy) model \(\boldsymbol{\pi}\) is trained by optimizing the following objective function:  
\[
\mathcal{L}_{\pi}(\theta_{\pi}) = \mathbb{E}_t \left[ \min \left( \rho_t(\boldsymbol{\pi}) \hat{A}_t, \text{clip}(\rho_t(\boldsymbol{\pi}), 1-\epsilon, 1+\epsilon) \hat{A}_t \right) \right],
\]
where \( \rho_t(\boldsymbol{\pi}) \) represents the importance sampling ratio between the current and previous policies, and \( \hat{A}_t \) is the advantage function, calculated using Generalized Advantage Estimation (GAE). The Critic (value) model \( V \) is trained by minimizing the following value loss function:
\begin{align}
\mathcal{L}_{\mathcal{V}}(\phi_V) = \frac{1}{N} \sum_{t=1}^N \left( V(s_t) - \hat{R}_t \right)^2,
\end{align}
where \( \hat{R}_t \) is the target return for time step \( t \).

\subsubsection{Double Experience Replay Buffer}
Overfitting and discrepancies in sample distributions across batches can cause the world model to enter abnormal iteration phases, where its predictions deviate significantly from true trajectory distributions. The pseudo trajectories generated during these phases, particularly reward samples~\cite{Wang_Liu_Li_2020}, can mislead the policy network, driving optimization in conflicting directions and disrupting convergence.

\begin{figure}[htbp]
	\centering
	\includegraphics[width=\linewidth]{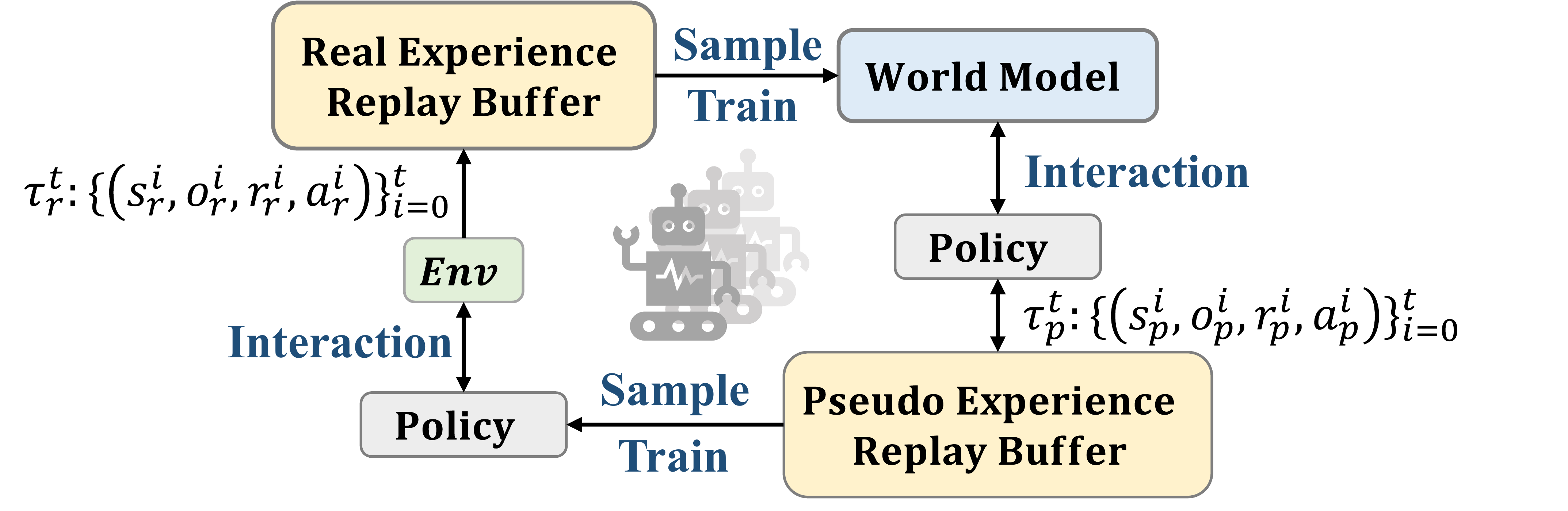}
	\caption{Dual Experience Replay Buffer structure.}
	\label{fig:drb}
\end{figure}

To address this, GAWM adopts a dual experience replay buffer structure, as shown in Fig.~\ref{fig:drb}. Alongside the original buffer for true trajectories, a pseudo trajectory buffer is introduced to reduce sample correlation and stabilize target distributions during training. Unlike training on single trajectory fragments, the dual buffer aggregates samples from multiple trajectories, enhancing diversity, mitigating overfitting, and improving the policy's generalization.
\subsection{Overall Algorithm Process}
\label{sec:overall}
\begin{algorithm}[htbp]
\caption{The training process of GAWM}
\label{alg:train}
	Initialize joint policy $\boldsymbol{\pi}$, world model $\mathcal{M}$, fusion block $\mathcal{F}$, real trajectory replay buffer $\mathcal{B}_r$, and pseudo trajectory replay buffer $\mathcal{B}_p$;

		\For{$N$ episodes}{
		Collect an episode of real-environment trajectory and add it to $\mathcal{B}_r$;
			
		\label{wm}\For(\tcp*[f]{\bluecomment{Train world model $\mathcal{M}$}}){$E_{\mathcal{M}}$ epochs}{ 
			Initialize $\boldsymbol{z}_t$ and $\boldsymbol{h}_t$;	Sample $\tau_r= \langle \boldsymbol{o}_t,\boldsymbol{a}_t,r_t,\boldsymbol{\gamma}_t,\boldsymbol{o}_{t+1} \rangle$ from $\mathcal{B}_r$;
			
			Use $\mathcal{M}$ for one-step temporal prediction and reconstruction on $\tau_r$;
			
			Calculate the joint one-step loss: $\mathcal{L}_\mathcal{M}(\theta_{\mathcal{M}}) = \mathcal{L}_{rec} + \beta \mathcal{L}_{\mathcal{KL}}$;
			
			Minimize $\mathcal{L}_\mathcal{M}(\theta_{\mathcal{M}})$ by gradient descent and update $\mathcal{M}$;
		}\label{wmend}

		\For(\tcp*[f]{\bluecomment{Train policy model $\pi$}}){$E_\pi$ epochs}{\label{pi}
			Initialize $\boldsymbol{z}_t$ and $\boldsymbol{h}_t$; Sample $\boldsymbol{o}_t$ from $\mathcal{B}_r$ as the initial data;
			
			\For{$k$ rollout steps}{
				Agents take action $\boldsymbol{a}_t$ according to $\boldsymbol{\pi}(\boldsymbol{a}_t|\boldsymbol{o}_t)$;
				
				$\mathcal{M}$ predicts $\{\boldsymbol{o}_{t+1}, r_{t+1}, \boldsymbol{\gamma}_{t+1}\}$ and stores them in $\mathcal{B}_p$;
				
				Let $\boldsymbol{o}_{t+1} = \boldsymbol{o}_t$, $t = t+1$;
			}
			\For{$E_{sample}$ epochs}{
				Sample $\tau_p=\langle \boldsymbol{o}_t,\boldsymbol{a}_t,r_t,\boldsymbol{\gamma}_t \rangle$ from $\mathcal{B}_p$;
				
				Compute $\boldsymbol{A}_t$ and returns on $\tau_p$ and compute $\mathcal{L}_{\boldsymbol{\pi}}(\theta_{\pi})$, $\mathcal{L}_{\mathcal{V}}(\phi_V)$;
				
				Minimize $\mathcal{L}_{\boldsymbol{\pi}}, \mathcal{L}_{\mathcal{V}}$ by gradient descent and softly update $\boldsymbol{\pi}(\boldsymbol{a}_t|\boldsymbol{s}_t),V(\boldsymbol{s}_t)$;
			}
		}\label{endpi}
	}
\end{algorithm}
As outlined in Algorithm~\ref{alg:train}, the training process consists of two key components: training the world model \(\mathcal{M}\) (lines~\ref{wm}–\ref{wmend} in Algorithm~\ref{alg:train}) and training the policy \(\boldsymbol{\pi}\) (lines~\ref{pi}–\ref{endpi} in Algorithm~\ref{alg:train}). For \(\mathcal{M}\), samples are drawn from the real experience replay buffer \(\mathcal{B}_r\), and \(\mathcal{M}\) is updated by minimizing a joint loss function comprising single-step temporal prediction and state reconstruction, using gradient descent (see Eq.~\ref{align:jointlossfunc}). During the policy training phase, \(\mathcal{M}\) is utilized to generate pseudo-sample trajectories, which are stored in the pseudo experience replay buffer \(\mathcal{B}_p\). Trajectories are then sampled from \(\mathcal{B}_p\), the policy advantage function and cumulative return are computed for these trajectories, and \(\boldsymbol{\pi}\) is updated using a soft update mechanism.
\section{Experiments}
In this section, we will present GAWM's empirical evaluation on multi-agent benchmarks. In Sec.~\ref{sec:asymptotic}, several baseline MARL methods will be compared with GAWM in SMAC benchmark. 
\paragraph{Environments} The Starcraft Multi-Agent Challenge     (SMAC)~\cite{Samvelyan_Rashid_Witt_Farquhar_Nardelli_Rudner_Hung_Torr_Foerster_Whiteson_2019} is a multi-agent discrete and collaborative control benchmark based on StarcraftII. Each task contains a scenario where there are two opposing teams, one controlled by the game robot and the other controlled by our algorithm. The goal is to defeat all the enemy agents. Our method and other baselines are tested on 8 maps of SMAC from \textit{easy} to \textit{super hard}, including \textit{2s\_vs\_1sc, 3s\_vs\_3z, 2s3z, 3s\_vs\_4z, 3s\_vs\_5z, 1c3s5z, 8m, corridor}.
\paragraph{Baselines} 
We compare GAWM with model-based and model-free baseline methods to assess the convergence performance of our approach in standard secenarios. The model-based methods include 1) MAMBA, 2) MAG. Model-free methods include 1) MAPPO~\cite{yu2022surprising}, 2) QMIX~\cite{rashid2020monotonic}.
\subsection{Performance Comparison}
\label{sec:asymptotic}
Now we will compare GAWM with other baselines in the SMAC environment. We assign three completely random seeds to each algorithm and conduct independent experiments to investigate the stationarity of the convergence process and the final convergence performance. After a fixed number of training steps, we saved the weight files of different algorithms and seeds and independently tested them for 1000 rounds to obtain the final convergence performance test results.
\begin{figure*}[t]
	\centering
	\includegraphics[width=\linewidth]{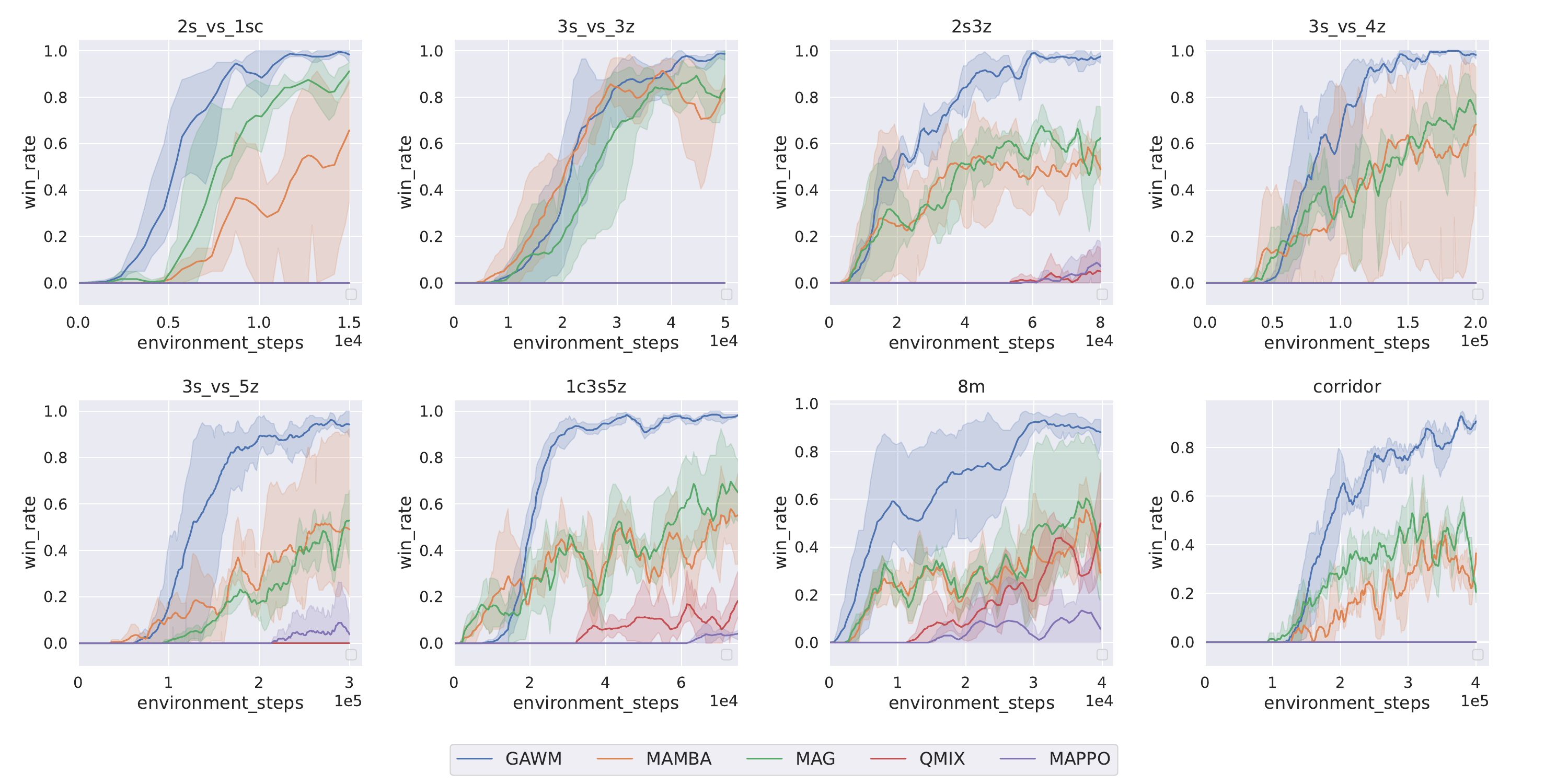}
	\caption{Comparisons with other baselines. The solid line represents the running average of 3 different random seeds, and the shaded area corresponds to the winning rate/episode rewards range for different seeds at the same time. The X-axis represents the number of steps taken in the real environment, and the Y-axis represents the win rate (SMAC).}
	\label{fig:result}
\end{figure*}
\begin{table}[h]
	\centering
	\setlength{\tabcolsep}{1mm}{%
		\begin{tabular}{c|cccccc}
			\hline
			Maps		      & GAWM       	     & MAG    & MAMBA   &MAPPO &QMIX      \\ \hline
			2s\_vs\_1sc(15k)  & \textbf{93(3)}   & 86(4)  & 64(15)  &0(0)  &0(0)   			  \\
			3s\_vs\_3z(50k)   & \textbf{95(3)}   & 83(6)  & 78(10)  &0(0)  &0(0)                  \\
			2s3z(80k)         & \textbf{98(1)}   & 67(11) & 71(12)  &9(2)  &3(1)   \\
			3s\_vs\_4z(200k)  & \textbf{97(1)}   & 81(11) & 64(32)  &0(0)  &0(0)      \\
			3s\_vs\_5z(300k)  & \textbf{93(2)}   & 55(12) & 53(8)   &8(1)  &0(0)      \\
			1c3s5z(75k)       & \textbf{98(3)}   & 65(13) & 54(9)   &15(3) &4(1)      \\    
			8m(40k)           & \textbf{90(2)}   & 63(8)  & 37(7)   &38(5) &12(3)     \\
			corridor(400k)    & \textbf{86(3)}   & 27(7)  & 39(9)   &0(0)  &0(0)      \\ \hline
			 
		\end{tabular}%
	}
	\caption{During the training process, the maximum episode steps (MES) is fixed for each map and scene. 
		After completing training for a specified number of real environment interaction steps (REIS) in different environments, the model weights are saved, and the average win rate (in SMAC) or episode reward (in MaMuJoCo), along with their standard deviations, are independently evaluated over 1000 test episodes. Bold numbers highlight the highest average performance among all baselines. GAWM consistently achieves the best performance across all tests. }
	\label{tab:experiments}
\end{table}

The comprehensive experimental results unequivocally demonstrate the superiority of our proposed approach, GAWM, over both model-based and model-free methods across all test maps and scenarios, even within a constrained number of iterations. As detailed in Tab.~\ref{tab:experiments}, GAWM, as a CTDE-based method, consistently outperforms other model-based MARL baselines (CTCE) and model-free MARL baselines (CTDE), achieving significantly higher performance metrics, exemplified by the win rate in SMAC. This superior performance underscores GAWM's exceptional convergence efficiency during training. Moreover, as illustrated in Fig.~\ref{fig:result}, the shaded regions in the training curves, representing the range between maximum and minimum win rates or episode rewards across different random seeds at each training step, are notably smaller for GAWM. This suggests that GAWM not only converges more effectively but also achieves greater consistency across random initializations. By leveraging a more centralized and robust global state representation structure, the world model of GAWM generates data samples with superior global coherence. This design ensures that the generated samples do not inadvertently shift towards divergent gradient directions, thereby maintaining optimization stability. The advantages of GAWM are evident across a variety of challenging scenarios. Notably, on maps with high action precision requirements, such as \textit{3s\_vs\_5z}, and on maps where the complexity of world model construction is significant, such as \textit{1c3s5z}, GAWM achieves remarkably superior performance. Even on the highly challenging \textit{corridor} map, which demands both precise action execution and sophisticated environmental modeling, GAWM consistently maintains optimal performance, highlighting its robustness and adaptability in diverse environments. As the complexity of the test scenarios increases, the benefits of GAWM become even more pronounced. The method demonstrates significant improvements in both sample efficiency and overall performance under more demanding conditions. These results emphasize GAWM's enhanced stability and robustness across a wide range of complex environments.We attribute these outstanding results to the synergy of several innovative strategies embedded in our approach. Notably, the enhancement of global information representation during observation fusion, coupled with advanced trend modeling mechanisms, plays a pivotal role in boosting GAWM's ability to adapt and excel in multi-agent reinforcement learning tasks.

\subsection{Ablation Studies}
\begin{figure*}[htbp]
	\centering
	\begin{minipage}{\linewidth}
		\centering
		\includegraphics[width=\linewidth]{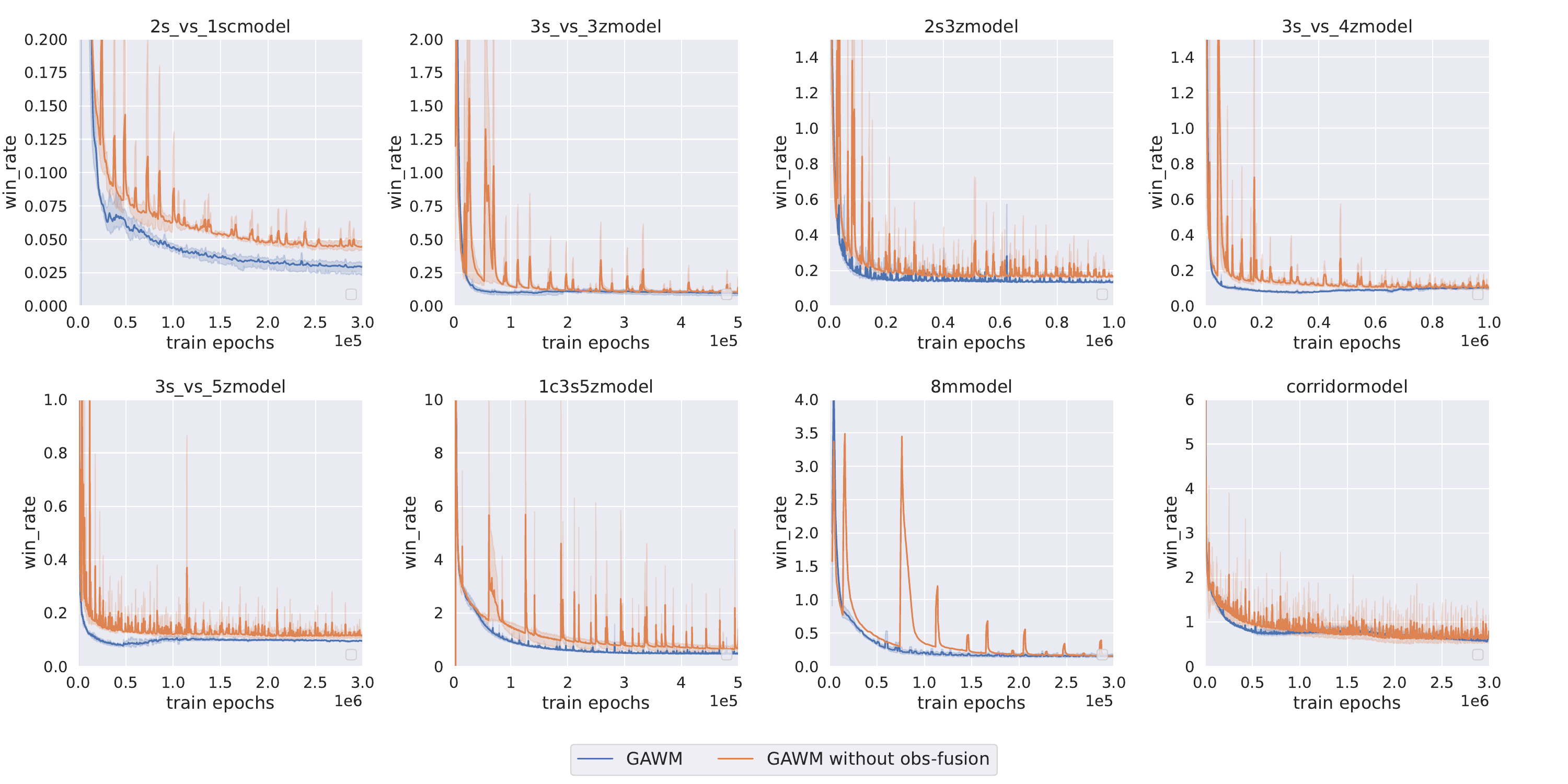}
		\caption{Training loss curve for the world model. The solid line represents the running average of 3 different random seeds, and the shaded area corresponds to the loss range for different seeds at the same time. The X-axis represents the number of training epochs of world model, and the Y-axis represents the loss value.}
		\label{fig:lossablation}
	\end{minipage}
	\hfill
	\begin{minipage}{\linewidth}
		\centering
		\includegraphics[width=\linewidth]{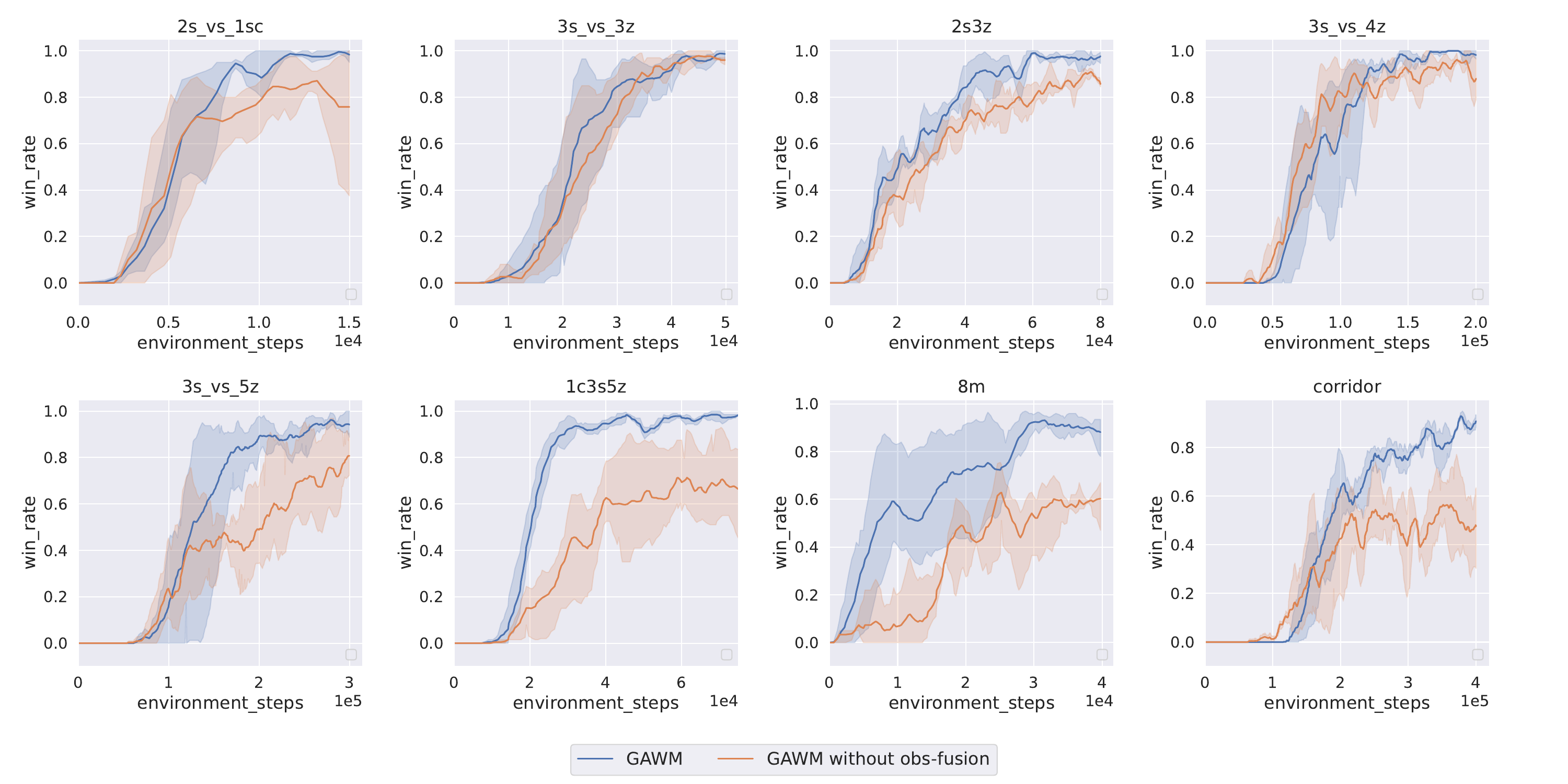}
		\caption{Win rate curve for ablation experiments. The solid line represents the running average of 3 different random seeds, and the shaded area corresponds to the winning rate range for different seeds at the same time. The X-axis represents the number of steps taken in the real environment, and the Y-axis represents the win rate (SMAC).}
		\label{fig:winrateablation}
	\end{minipage}
	\label{fig:combined_results}
\end{figure*}

We conducted a targeted ablation study to validate the effectiveness of our method in enhancing the robustness of the world model training process. Specifically, as depicted in Fig.~\ref{fig:lossablation} and Fig.~\ref{fig:winrateablation}, we compared the loss function curve and the real-time win rate curve of the world model between the full GAWM framework and a variant of GAWM without the observation fusion (obs-fusion) module. The results reveal several critical insights into the role of obs-fusion in stabilizing training dynamics and improving performance.

The experimental results clearly demonstrate that removing the obs-fusion module leads to substantial instability in the world model training process. Without obs-fusion, the posterior model directly relies on distributed observation information for state reconstruction, which introduces frequent and pronounced fluctuations in the loss function across nearly all test maps, as shown in Fig.~\ref{fig:lossablation}. This instability in the loss function translates to a highly unstable distribution of generated pseudo data samples. Such instability adversely affects the training dynamics, leading to significant non-stationarity in policy convergence. In contrast, incorporating the obs-fusion module and global state predictors yields notable improvements in training stability and performance. As illustrated in Fig.~\ref{fig:winrateablation}, these components significantly mitigate policy fluctuations and enhance sample efficiency, enabling the model to converge more effectively and consistently. The benefits of this approach are particularly pronounced in more complex scenarios, such as \textit{3s\_vs\_5z}, \textit{1c3s5z}, and \textit{corridor}, which demand precise action coordination and robust state representation due to their higher complexity and dynamic nature. By enabling a more centralized and coherent global state representation, the obs-fusion module ensures that the posterior model processes more stable and globally consistent information, which in turn stabilizes the training dynamics of the world model. This improvement not only reduces the variance in generated data samples but also facilitates smoother policy updates, ultimately leading to superior overall performance. These findings underscore the critical role of obs-fusion and global state prediction mechanisms in addressing the challenges of multi-agent reinforcement learning, particularly in scenarios that involve complex interactions and high-dimensional state spaces.

\subsection{Model Analysis}
\label{sec:gci_analysis}
Due to the fact that MBRL is essentially an online learning process, there is a lack of validation steps to verify the enhancement effect of GAWM on data generation. Therefore, we designed an independent offline testing phase to verify the superiority of GAWM's world model. To rigorously evaluate the performance of the world model in generating globally consistent and accurate multi-agent data samples, we introduce the Global Consistency Index (GCI) and an accuracy metric, termed Global Prediction Error (GPE). These metrics assess (1) the degree of consistency in predicted observations and shared environment variables among agents and (2) the accuracy of the predictions relative to the true value.  

\subsubsection{Metrics Definition}
\noindent\textbf{Global Consistency Index (GCI):}  
The GCI quantifies conflicts in the predicted global state representations, rewards, and discount factors among agents. Each agent predicts a local global state, \(s_t^i\), which is derived from its observation, \(o_t^i\), at time \(t\), and includes the visible environmental information and the states of opponent agents. The mean global state, \(\bar{s}_t\), is computed across all agents. The GCI measures the inconsistency between agents by comparing their predictions of the global state, rewards, and discount factors. A higher GCI indicates greater inconsistency. The GCI is calculated as:
\begin{equation}
\text{GCI} = \frac{1}{T} \sum_{t=1}^{T} \frac{1}{N} \sum_{i=1}^{N} \Big( \| \hat{s}_t^i - \bar{s}_t \|_2 + \mathbb{I}(|\hat{r}_t^i - \bar{r}_t| > \epsilon_r) + \mathbb{I}(|\hat{\gamma}_t^i - \bar{\gamma}_t| > \epsilon_\gamma) \Big),
\end{equation}
where \(T\) is the total number of time steps, \(N\) is the number of agents, \(\hat{s}_t^i\) is the local global state predicted by agent \(i\) at time \(t\), which is derived from its observation, \(o_t^i\), \(\bar{s}_t\) is the mean global state across all agents at time \(t\), \(\hat{r}_t^i\) and \(\bar{r}_t\) are the predicted and mean rewards at time \(t\), \(\hat{\gamma}_t^i\) and \(\bar{\gamma}_t\) are the predicted and mean discount factors at time \(t\), \(\mathbb{I}(\cdot)\) is the indicator function, and \(\epsilon_r\), \(\epsilon_\gamma\) are thresholds for acceptable deviations. A low GCI reflects greater consistency among agents, implying a more reliable global representation.

\noindent\textbf{Global Prediction Error (GPE):}  
The GPE evaluates the accuracy of the world model's predictions by comparing them to the true value. It is defined as:
\begin{equation}
\text{GPE} = \frac{1}{T} \sum_{t=1}^{T} \frac{1}{N} \sum_{i=1}^{N} \left( \| \hat{o}_t^i - o_t^i \|_2 + |\hat{r}_t^i - r_t| + |\hat{\gamma}_t^i - \gamma_t| \right),
\end{equation}
where $o_t^i$, $r_t$, and $\gamma_t$ are the true observation, reward, and discount factor at time $t$. A lower GPE indicates better predictive accuracy of the world model.

\subsubsection{Experimental Design}
After training with a fixed number of steps, we extracted the model weights and conducted separate tests. We evaluate GAWM, GAWM without obs-fusion (GAWM*), and other baseline methods (e.g., QMIX, MAMBA) on four maps: \textit{3s\_vs\_5z}, \textit{8m}, \textit{1c3s5z}, and \textit{corridor}. Each method generates 1000 pairs of pseudo trajectory segments and real trajectory segments, and we conduct offline testing on these data pairs. These data pairs are uniformly sampled from the entire time series to ensure that the sampled segments cover the entire convergence process evenly.

\subsubsection{Experimental Result}

\begin{table}[htbp]
	\centering
	\begin{tabular}{@{}c|cccc@{}}
		\hline
		Map & GAWM & GAWM* & MAG & MAMBA \\ \hline
		\textit{3s\_vs\_5z}    & \textbf{0.63$\pm$0.02} & 1.98$\pm$0.10 & 2.43$\pm$0.12 & 2.91$\pm$0.14  \\ 
		\textit{1c3s5z}        & \textbf{1.23$\pm$0.05} & 2.73$\pm$0.15 & 3.48$\pm$0.18 & 4.14$\pm$0.20  \\ 
		\textit{8m}      	   & \textbf{0.75$\pm$0.03} & 2.22$\pm$0.11 & 2.46$\pm$0.13 & 3.54$\pm$0.17  \\ 
		\textit{corridor}      & \textbf{1.29$\pm$0.06} & 2.52$\pm$0.14 & 3.03$\pm$0.15 & 3.87$\pm$0.18  \\ \hline
	\end{tabular}
	\caption{Comparison of Global Consistency Index (GCI) across different methods and scenarios. Lower values indicate better performance.}
	\label{tab:gci_results}
\end{table}

\begin{table}[htbp]
	\centering
	\begin{tabular}{@{}c|cccc@{}}
		\hline
		Map 				   & GAWM            & GAWM*  & MAG    & MAMBA  \\ \hline
		\textit{3s\_vs\_5z}    & \textbf{0.70$\pm$0.03} & 0.99$\pm$0.09  & 1.35$\pm$0.17 & 1.69$\pm$0.28  \\ 
		\textit{1c3s5z}        & \textbf{1.28$\pm$0.06} & 1.72$\pm$0.11  & 2.05$\pm$0.21 & 2.37$\pm$0.24  \\ 
		\textit{8m}      	   & \textbf{0.22$\pm$0.01} & 0.43$\pm$0.12  & 0.58$\pm$0.15 & 0.65$\pm$0.23  \\ 
		\textit{corridor}      & \textbf{1.33$\pm$0.05} & 1.77$\pm$0.19  & 2.19$\pm$0.21 & 2.80$\pm$0.25  \\ \hline
	\end{tabular}
	\caption{Comparison of Global Prediction Error (GPE) across different methods and scenarios. Lower values indicate better performance.}
	\label{tab:gpa_results}
\end{table}
GAWM consistently outperforms the other methods across all evaluation metrics, with lower GCI and GPE values. In addition to its superior performance, GAWM demonstrates greater stability, as evidenced by its smaller standard deviations compared to methods like GAWM* (without observation fusion), MAG, and MAMBA. The larger standard deviations observed in the baseline methods indicate higher variability in both consistency and accuracy, suggesting that they are less robust and exhibit more fluctuations across different trials. This makes GAWM not only more effective but also more reliable, maintaining stable performance across a range of scenarios.
\section{Conclusion}
In this article, we introduce GAWM, a model-based multi-agent reinforcement learning (MARL) algorithm that significantly enhances the global state representation capabilities of the RSSM-structured world model. This is achieved by incorporating a state reconstruction architecture and trend modeling with global information fusion via Transformer mechanisms. As a result, GAWM markedly improves both the global consistency and the stability of the data distribution in the generated data samples. Within a fixed number of training steps, GAWM outperforms state-of-the-art model-free and model-based methods in terms of sample efficiency, strategy performance, and stability. By further optimizing the multi-agent world model within the RSSM framework, GAWM paves the way for more effective applications of model-based reinforcement learning (MBRL) in complex multi-agent environments. However, GAWM does have some limitations. For example, the inclusion of additional Transformer components slightly increases the per-iteration training time of the world model.

\bibliographystyle{elsarticle-num} 
\bibliography{references.bib}


\clearpage
\end{document}